\documentclass[twocolumn,showpacs,preprintnumbers,amsmath,amssymb]{revtex4}

\usepackage{graphicx}
\usepackage{dcolumn}
\usepackage{bm}

\begin{document}

\title{Correlations and Scaling Laws in Human Mobility}

\author{Xiang-Wen Wang$^{1,2}$}
\author{Xiao-Pu Han$^{3}$}
\author{Bing-Hong Wang$^{2,4,5}$}

\affiliation{
$^{1}$Department of Physics, Virginia Polytechnic Institute and State University, Blacksburg, Virginia 24061-0435, USA\\
$^{2}$Department of Modern Physics, University of Science and Technology of China, Hefei 230026, China\\
$^{3}$Institute of Information Economy and Alibaba Business College, Hangzhou Normal University, Hangzhou 310036, China\\
$^{4}$College of Physics and Electronic Information Engineering, Wenzhou University, Wenzhou 325035, China\\
$^{5}$The Research Center for Complex System Science, University of Shanghai for Science and Technology, Shanghai, 200093 China
}

\date{\today}

\begin{abstract}

Human mobility patterns deeply affect the dynamics of many social systems. In this paper, we empirically analyze the real-world human movements based GPS records, and observe rich scaling properties in the temporal-spatial patterns as well as an abnormal transition in the speed-displacement patterns. We notice that the displacements at the population level show significant positive correlation, indicating a cascade-like nature in human movements. Furthermore, our analysis at the individual level finds that the displacement distributions of users with strong correlation of displacements are closer to power laws, implying a relationship between the positive correlation of the series of displacements and the form of an individual's displacement distribution. These findings from our empirical analysis show a factor directly relevant to the origin of the scaling properties in human mobility.
\end{abstract}


\maketitle

\section{Introduction}

The statistical patterns of human daily movements directly affect the physical contacts between humans and thus deeply impact the dynamics of many social systems. 
The understanding of real-world human mobility patterns would be much helpful for many aspects in social dynamics, such as epidemics spreading \cite{Belik, Balcan,WangL,Ni}, the designing of traffic systems \cite{Horner}, or localized recommendations \cite{Clements,Scellato}. Since the pioneering work of  Brockmann et al \cite{7}, the  temporal-spatial statistical properties in human movements have become a new issue in complex sciences and have attracted much attention in recent years.

The most dramatic discovery in the statistical patterns of human mobility is the existence of wide-spread scaling properties \cite{7, Gonzalez, Song1}.
The first one is the power-law-like displacement distribution, which has been observed in many empirical analyses of real-world human movements \cite{7, Rhee, Gonzalez} ageographicnd even in the virtual world of online-games \cite{Szell}. This result sharply differs with the traditional understanding based on random walks, and reveals long-range correlations in human travels and social interactions. Other scaling properties include the staying time distributions which denote that humans usually stay in a few locations quite a long time \cite{Gonzalez},
and the visitation frequency distributions are dominated by a few locations that are usually much more frequently visited \cite{Gonzalez, Song1}, and so on.

Many other abnormal properties are also found in human mobility patterns, including ultra-slow diffusion\cite{7,Gonzalez}, anisotropism \cite{Gonzalez}, high predictability \cite{Song}, and the limitation of roads \cite{Jiang2009}.
These discoveries reveal abnormal features in real-world human mobility, in stark contrast to the traditional understandings based on the hypothesis of random-walk-like human mobility or on that of L\'evy flights with the same scaling displacement distributions.

However, these findings are still facing several controversies. Due to the limitation of original data, most previous works are at the population level, and a direct analysis of individuals is rarely seen. Recently, Yan, et al. \cite{Yan2} reported the diversity in individual-level mobilities and found out that most of the individuals' displacement distributions do not obey the scaling law.
Moreover, several recent researches indicated that the move length in human urban trips or the travels by a single type of transportation do not obey well a power law \cite{Baz1, Liang, Noulas}. These controversies require the confirmation from a more in-depth empirical analysis of human mobility patterns.

Recent studies also proposed many models to explain the underlying mechanisms that drives the emergence of these anomalies in human mobility. Generally, the basic dynamics of previous modeling works can be divided into the following classes: i) The descriptive models: L\'evy flights \cite{Rhee}, Self-similar least action walk (SLAW) \cite{Lee}, and Continuous-time random walks \cite{7};
ii) The exploration of new locations and the preferential return to visited places \cite{Song1}; iii) The effect of hierarchical traffic systems \cite{Han}; iv) the effect of few dominant trips \cite{Yan1}; v) The spatial heterogeneity of population density or the geographic locations \cite{Noulas, Vene}; vi) The radiation model proposed by Simini et. al. \cite{Simini}, which can reproduce many mobility patterns at the global level; vii) The aggregation of individuals without scaling properties \cite{Yan2}.
These models can reproduce parts of the empirical findings. Nevertheless, it is difficult to identify common rules from these model, and thus it remains controversial what drives the emergence of these abnormal properties in human mobility. It would therefore be helpful if the empirical analysis can identify characteristic factors affecting the emergence of these anomalies.

In this paper, based on the empirical analysis of GPS data sets, we report one of the characteristic factors that is relevant to the scaling displacement distributions: the correlation of the series of displacement.
We first show the aggregated temporal-spatial properties at the population level (Section II) and then we analyze the correlation of the aggregated series of displacements (Section III). Finally, we discuss the diversity in individuals' mobility patterns and the relationship between the correlation of the series of displacements and the scaling properties of displacement distributions (Section IV). We show that the correlation is indeed a tool that allows to investigate the underlying mechanisms from the empirical data.

\section{The scaling properties at the population level}

The data set in our analysis contains records from 165 volunteers that have been gathered over three years (April, 2007 -- Sep., 2010). The GPS trajectories result from the Microsoft Research Asia in Geo-life Project \cite{11,12,13}. More details can be found in {\it Appendix A}.

\begin{figure}
  \includegraphics[width=8.0cm]{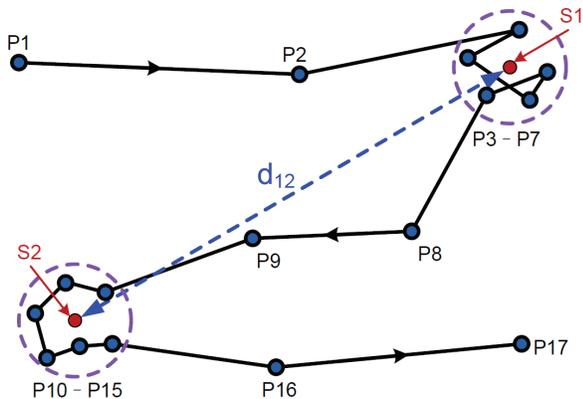}\\
  \caption{(Color online) Identification of two different staying points. P1-P17 represent 17 track points recorded by a GPS equipment from which we obtain two staying points S1 and S2. The displacement of travel is defined as the distance between the centers of the two staying points.  }\label{jud}
\end{figure}

\begin{figure*}
  \includegraphics[width=16.5cm]{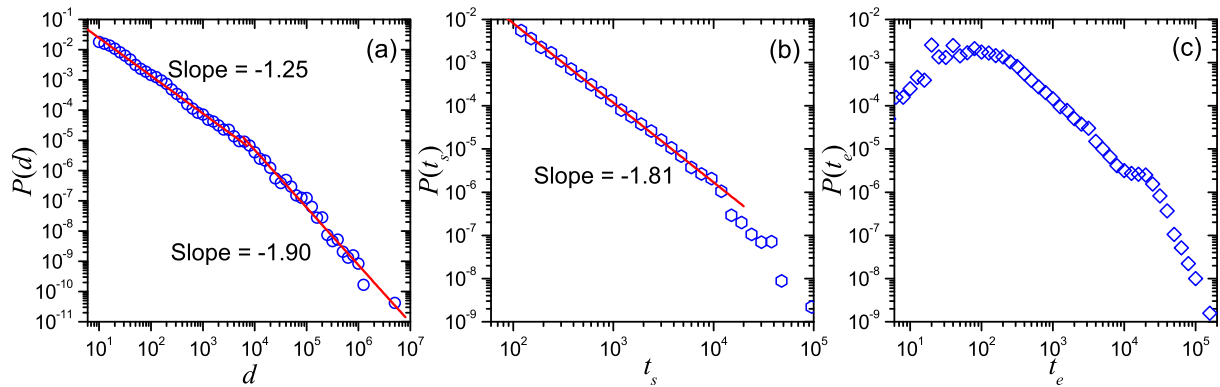}\\
  \caption{(Color online) (a) The aggregated displacement distribution $P(d)$,(b) the staying time distribution $P(t_s)$, (c) and the elapsed time distribution $P(t_e)$ in log-log plots.}\label{PLD}
\end{figure*}

\begin{figure}
  \includegraphics[width=8.7cm]{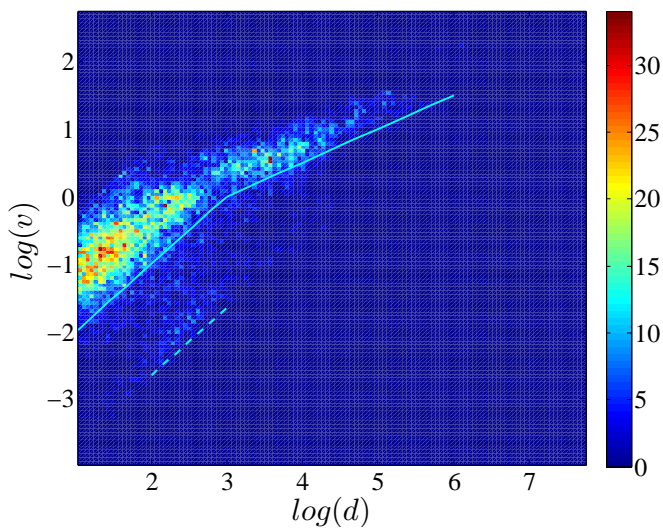}\\
  \caption{(Color online) The relationship between the average speed $v$ and the displacement $d$, where the slopes of the upper lines are 1.0 and 0.5 respectively, whereas the slope of the lower dashed line is 1.0.
  }\label{speed}
\end{figure}

We determine the effective staying positions from the dataset using a resolution of 10 meters in space and 120 seconds in time. Fig. \ref{jud} illustrates for a case of two staying positions, S1 and S2, are obtained from a sequence of GPS records. Details of our approach can be found in \emph{Appendix A}.
The geographical distance between two consecutive staying positions, e.g. S1 and S2 in Fig. \ref{jud}, is defined as the displacement of travel. The staying time in each staying position is defined as the time interval between the first and last GPS records in the given staying position.

Using the above method, we obtain 927 trajectories with recording times longer than 6 hours that contain 19376 effective staying points. The total staying time is 4463 hours, and the total displacement is 95472.33 kilometers.
From each of these trajectories, we can obtain a sequence that contains the staying positions, displacements and staying times.

We combine the displacements and staying times in all 927 files to calculate the displacement distribution and the staying time distribution at the population level. After log-binning, the displacement distribution $P(d)$ generally obeys the following power-law function with two different regimes (Fig.\ref{PLD} (a)):
\begin{equation}
    P(d) \sim \left\{
    \begin{array}{cc}
d^{-1.25}, &(d < 6.5km), \\
d^{-1.90},&(d \geq 6.5km).
    \end{array}
    \right.
\end{equation}
This power-law displacement distribution indicates that the typical behavior consists of many short-range trips and few long-range travels.
This conclusion is in substantial agreement with the conclusions of several previous findings \cite{Gonzalez,Song1}. The transition at $d \simeq 6.5$km is related to the average extend of the urban district of cities, indicating the difference between human urban movements and intercity travels. This difference may be due to the convenience of urban movements and the dominant high-frequency movements between few positions (such as home and working places) \cite{Yan1,Yan2}.

A similar scaling property is also observed in the staying time distribution $P(t_s)$ at the population level, which can be well fitted by a power-law function with exponent $-1.98$ (Fig. \ref{PLD} (b)), indicating that humans usually stay in few positions a quite long time. This result is also close to previous findings based on other data sets \cite{Gonzalez, Baz1, Baz2, Rhee}.

The distribution $P(t_e)$ of the elapsed time $t_e$ that individuals spend on the way from an effective staying position to the next one has also been studied. As shown in Fig. \ref{PLD} (c), $P(t_e)$ shows a strange behavior where two power-law-like sections are separated by an unusual bump when $10^4 < t_e < 2 \times 10^4$ seconds. It seems that this bump results from traffic jams. This result is somewhat different to the previous findings in urban taxi data \cite{Liang}.

Moreover, we calculate the average speed $v_i = d_i/t_{ei}$ for every user i, and plot each pair ($v_i$,$d_i$) on the plane to get the pattern of the relationship between speed and displacement. We surprisingly find that $v$ vs. $d$ generally obeys the bilinear form in a log-log plot, in which the first section ($d<10^3$ meters) is linear, whereas another part ($d>10^3$m) is sublinear (slope $\approx 0.5$), as shown in Fig. \ref{speed}. The point of transition $d \approx 10^3$ meters and $v \approx 1$ m/s, could relate to the length and speed of walking, therefore the two sections would correspond to the travel by foot or by automobile, with humans preferring a trip by automobile (bus, car, etc.) for distances longer than 1 kilometer.

In addition, some movements have ultraslow speed, as indicated by the dashed line in Fig. \ref{speed}. The corresponding displacements of these ultrashow movements are generally between $10^2$ meters and $10^3$ meters, and the corresponding elapsed time is mainly in the range from $10^4$ seconds to $2 \times 10^4$ seconds, corresponding to the bump in $P(t_e)$ displayed in Fig.\ref{PLD}, possibly indicating displacements hampered by traffic jams.


\section{Correlations of displacements at the population level}

\begin{figure*}
  \includegraphics[width=16cm]{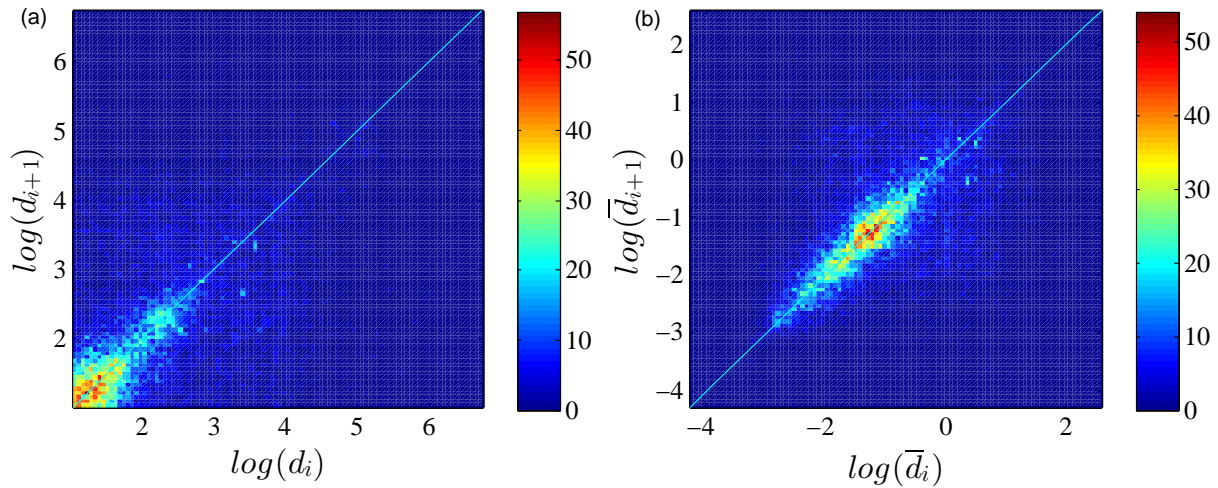}\\
  \caption{(Color online) Movement correlation scatter plot at the population level. Taking a displacement as abscissa, and the next displacement as ordinate, we get a scatter of $(d_i, d_{i+1})$ in a log-log plot. The figure shows a high density of points near the diagonal line $d_{i+1}=d_i$, which means a strong positive correlation between consecutive displacements.}\label{corr}
\end{figure*}

For each sequence of displacements of individuals, the correlation between two consecutive displacements reflects the trends and causal relationship in human travels.
To get the pattern of the correlation, we plot each of the data points $(d_i, d_{i+1})$ and calculate the density of these data points. Here $d_i$ and $d_{i+1}$ denote the $i$-th and the $(i+1)$-th displacement in the series $\vec{d}$. As shown in Fig. \ref{corr}(a), most of the data points $(d_i, d_{i+1})$ accumulate close to the diagonal line $d_{i+1} = d_i$, corresponding to a positive correlation.

We also plot the pattern using the related displacement $\bar{d}= d/d_*$, where $d_*$ is the average displacement of the user. We first calculate the average displacement of each user and then obtain the sequences of $\bar{d}$ from each file. Fig. \ref{corr}(b) shows the pattern of the density of the data points $(\bar{d_i}, \bar{d}_{i+1})$, where the positive correlation is much clearer.

Taking into account the heterogeneous $d$, we use the rank-based correlation coefficient named {\it Kendall's Tau} to quantify the strength of this correlation.
We first set $\vec{d}_i$ = $\{d_1, d_2, \cdots, d_i, \cdots, d_{N-1}\}$, and $\vec{d}_{i+1}$ = $\{d_2, d_3, \cdots, d_{i+1}, \cdots, d_N\}$  for every series, where $N$ is the total number of the displacements in the corresponding file.
The detailed introduction of Kendall's Tau can be found in {\it Appendix B}. The value of the Kendall's Tau $\tau_K = 0.424$ for the series $\vec{d}_i$ and $\vec{d}_{i+1}$, and the confidence interval with 95\% significance level is $0.015$. For the related displacement series $\vec{\bar{d}}_i$ and $\vec{\bar{d}}_{i+1}$, $\tau_K = 0.435$ which represents a significant positive correlation.


This remarkable positive correlation shows that a trip can have effect on the next one: if the current displacement is long, the next one has a high probability to be only slightly different. The change in displacement is usually gradual.
This gradual change agrees with our daily experience. For example, if we travel to another city, we first need to find a hotel in the city. The movement from our city to the target hotel generally is a long travel (the length may be several hundred miles). In the next several days, we might leave the hotel to visit some places around the city (generally tens of miles). During each trip, our visit will contain many short moves (usually less than one mile). A direct trip from our city to the place in the target city rarely appears.

\begin{figure}
  \includegraphics[width=8.7cm]{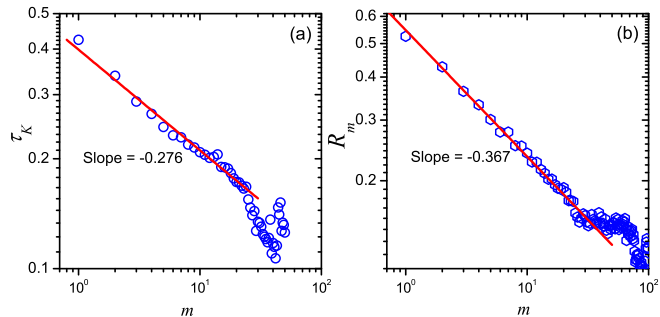}\\
  \caption{(Color online)
  (a) The decay of Kendall's Tau $\tau_K$ between $\vec{d}_i$ and $\vec{d}_{i+m}$ as a function of the interval $m$. (b) Pearson correlation coefficient $R_m$ between $log(\vec{\bar{d}}_i)$ and $log(\vec{\bar{d}}_{i+m})$ as a function of the interval $m$.
  }\label{Tau}
\end{figure}

Furthermore, to investigate the long-term correlations in human mobility, we calculate the Kendall's Tau $\tau_K$ of the series $\vec{d}_i$ and $\vec{d}_{i+m}$ ($m = 1, 2, \cdots $), and find that the function $\tau_K$ vs. $m$ shows a remarkable slow decay, which can be well fitted by a power-law function with an exponent $-0.276$ (Fig. \ref{Tau}(a)), implying that the effect of previous movements can continue a very long time. To ensure it, we plot the Pearson correlation coefficient $R_m$ between $log(\vec{\bar{d}}_i)$ and $log(\vec{\bar{d}}_{i+m})$. It does obey a power-law decay with a slope $\gamma = -0.367$ (Fig. \ref{Tau}(b)). The value $H = 1+\gamma/2 \simeq 0.82$ is the well-known Hurst exponent that denotes the long-term correlations in the fluctuation of the series \cite{Hurst}. Using the method of detrended fluctuation analysis (DFA) \cite{DFA} (see {\it Appendix C}), we also obtain a similar the Hurst exponent value $H = 0.87$, indicating a strong long-term correlation among the displacement series.

We also studied two other correlations, namely the correlation among the series of staying times, and the correlation between the staying time and the displacement. The series of staying times show only a weak positive correlation (its $\tau_K = 0.120$ with the confidence interval $0.015$), and the staying times and displacements are almost independent (its $\tau_K = 5.30 \times 10^{-3}$).

\section{Mobility patterns at the individual level}

\begin{figure*}
  \includegraphics[width=17cm]{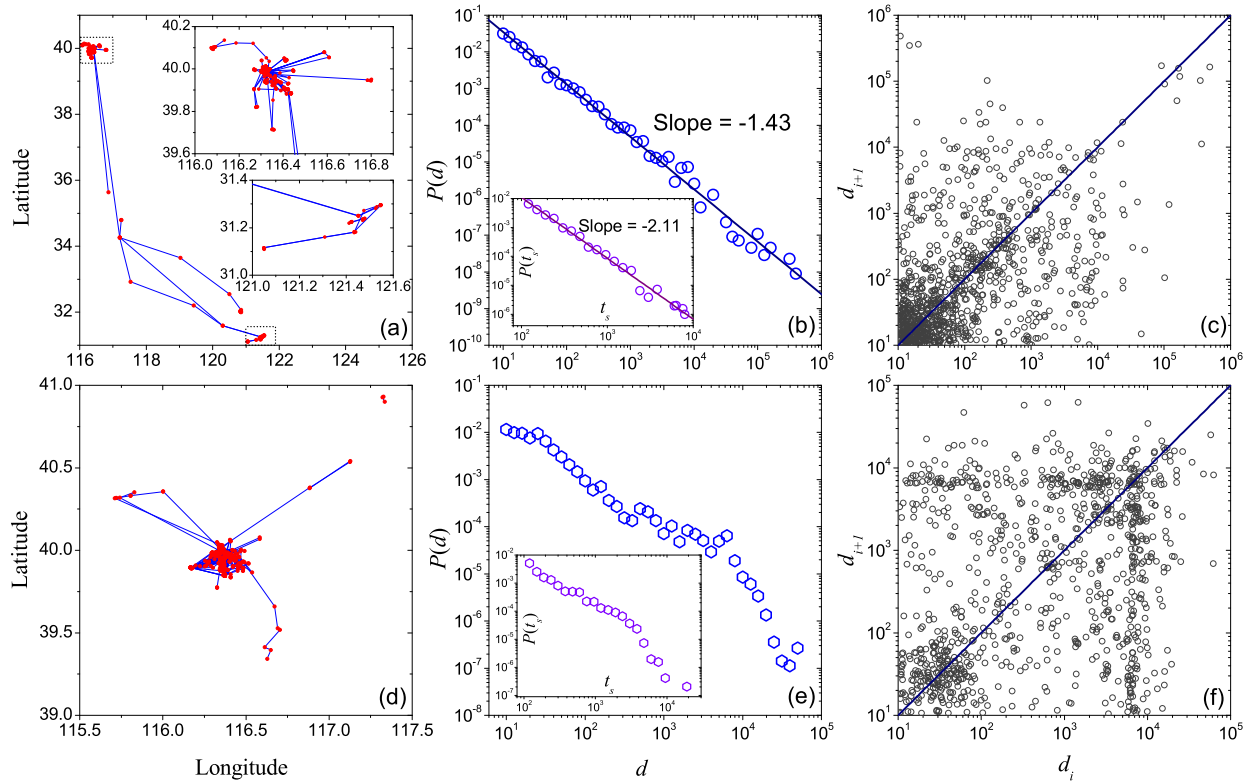}\\
  \caption{(Color online) Trajectories (a, d), displacement distributions $P(d)$ (b, e), staying time distributions $P(t_s)$ (insets in (b, e)) and correlation patterns (c, f) of two typical individuals (upper and lower three panels for individual No. 9 and No. 22 respectively}\label{9-22}
\end{figure*}

The above discussions showed the scaling patterns and positive correlation of human movements at the population level. Nevertheless, since the above results are aggregated over all individuals, we can not directly conclude that the movements of each individual also exhibit the same properties.
Actually, power-law-like displacement distribution at the population level can even be observed in a system where all the individuals' movements are Poissonian \cite{Petro, Yan2}.
Because of the lack of direct evidence, it remains controversial whether the scaling mobility patterns are universal at the individual level. Recently, Yan et al. reported the diversity of human mobility patterns at the individual level and that many individuals' displacement distributions usually are dominated by some frequently-appearing mobilities \cite{Yan2}. Due to the limitations in the original data sets of Yan's work, this conclusion still needs to be confirmed by more in-depth empirical studies based on datasets with higher resolution.

Among the 100 remaining users, we choose the users who had more than 200 effective staying positions to study their mobility patterns at the individual level, 200 being almost the lowest bound to obtained efficient statistical patterns. By doing this, 32 effective individuals with 698 files and 15189 staying positions are chosen. The number of effective staying positions $n_s$ and the number of displacements $n_d$ of each of the 32 users are listed in Table 1.

Plotting the displacement distribution $P(d)$ and correlation patterns $(d_i, d_{i+1})$ for each of the 32 users, we remark that users with stronger positive correlation seem to have usually a displacement distribution that is closer to a power law.
The trajectories, displacement distributions and correlation patterns $(d_i, d_{i+1})$ of two typical users are shown in Fig. \ref{9-22}. User No. 9 has many long-range movements, and his/her displacement distribution obeys well a power law. Significant positive correlation is also observed. In contrast, the displacement distribution of user No. 22 is bimodal-like, and the correlation is also not obvious.

The positive correlation reflects a gradually changing nature of human displacements. Previous studies in the temporal patterns have found that this gradually changing process, or say the cascading effect, is much relevant to the emergence of burstiness in human activities \cite{18}, as well as the long-term persistences \cite{Ryb1, Ryb2}.
Our results seem to indicate that the positive correlation in the displacements is related to the scaling properties in human mobility patterns.


To prove this hypothesis, we need to test the relationships between the strength of the correlation and the form of the users' displacement distributions.

Using the method introduced above, we first calculate the Kendall's Tau of the series $\vec{d}_i$ and $\vec{d}_{i+1}$ for each user, as shown in Table 1. Although all $\tau_K$ of the 32 users are positive, the value varies in a wide range from 0.2 to 0.5, showing a great diversity in the correlation. More than 2/3 of all users (23/32) have the Kendall's Tau $\tau_K>0.3$ and exhibit significant positive correlation.

The correlation coefficients $R_a$ of each user's displacement series are also calculated. Due to the heterogenous displacements, the logarithm of displacement $\ln d$ is used here, so $R_a$ is defined as:
\begin{equation}
R_a = \frac{\langle (\log(d_i) - \log(d_*))(\log(d_{i+1}) - \log(d_*))\rangle}{\sigma^2},
\end{equation}
where $d_*$ is the average displacement of the user and $\sigma$ is the variance of the displacement series $\vec{d}$. The values of $R_a$ for all 32 users are shown in Table 1. All of them are higher than 0.5, showing strong positive correlation in agreement with the above results for the method of Kendall's Tau.
And also, to quantify the long-term correlations, we calculate the Hurst exponent $H$ of the series $\log{\vec{d}}$ of each user using DFA and find $H>0.5$ for all of them (Table 1), showing significant long-term persistence on displacements. 

To check whether individual-level displacement distributions exhibit a power-law form, we plot these distributions and find that most of them seem to be power-law-like after log-binning. Here the Kolmogorov-Smirnov Test (KS Test) \cite{19} is used to test the power-law fits of these empirical data points. After estimating and setting a lower bound $x_{min}$ in the dataset, KS test will return confidence probability $p_{KS}$. Generally speaking, the bigger $p_{KS}$ is, the better the fit is. Table 1 shows $p_{KS}$ of the log-binning displacement distribution for each user, in which most of them have $p_{KS} > 0.1$ and have a well-fitted power-law-like section.

However, several users have very large estimated values for $x_{min}$ in the KS test, showing that the power-law-like section only covers a small range in the tail of $P(d)$. We therefore fix $x_{min}$ to 10 meters to test if $P(d)$ can be well fitted by a power law in all of the range. This yields the confidence probability $p'_{KS}$. Unfortunately, in only a few users is the requirement $p'_{KS} > 1$ fulfilled, as shown in Table 1, indicating that for most of these users a power law is observed over only a small range.

To quantify the differences between $P(d)$ and strict power law, one can also directly fit the data points of $P(d)$ to get the Pearson correlation coefficient $R'_d$ between the fitting curve and $P(d)$\cite{21}. The better fitting corresponds to smaller negative values of $R'_d$ due to the decaying power-law function, and $R_d = -1$ for the $P(d)$ that completely coincides with a power law. As shown in Table 1, all users' $R'_d$ are less than $-0.9$.

\begin{figure}
  \includegraphics[width=8.9cm]{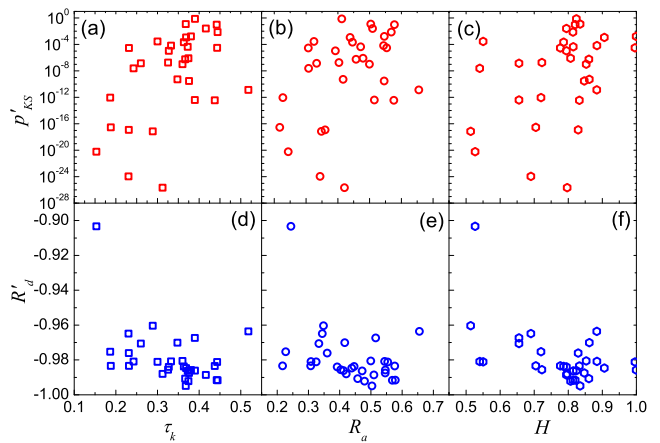}\\
  \caption{(Color online) Patterns show the correlations: (a) $p'_{KS}$ vs. $\tau_K$, (b) $p'_{KS}$ vs. $R_a$, (c) $p'_{KS}$ vs. $H$, (d) $R'_d$ vs. $\tau_K$, (e) $R'_d$ vs. $R_a$, (f) $R'_d$ vs. $H$. Kendall's Tau of these correlations respectively are (a) 0.367, (b) 0.246, (c) 0.278, (d) -0.274, (e) -0.286, (f) -0.254, with 95\% significance level confidence interval 0.262. }\label{indcorr}.
\end{figure}

Now we have five quantities for each individual, $\tau_K$, $R_a$ and $H$ for the correlations of user's displacements, $p_{KS}$ and $R'_d$ for the quality of the power-law fitting. We plot six relationships of these quantities and respectively calculate their Kendall's Tau values, as shown in Fig. \ref{indcorr}. Most of these correlations are significant, and in supporting of
our previous guess that the scaling mobility patterns usually correspond to higher correlation of displacements. This result implies that the cascading-like processes play an important role in the emergence of the scaling properties in human movements.

However, unlike the previous findings in human communications \cite{Ryb1,Ryb2}, the long-term correlations of move-lengths look independent of the power-law exponents of $P(d)$ (The Kendall's Tau between $\alpha$ and $H$ is $-0.129$ with 95\% significance level confidence interval $0.262$).

Similarly, we calculate the Pearson correlation coefficient $R'_t$ between the staying time distribution $P(t_s)$ and power law fits for each individual, as shown in Table 1.
However, $R'_t$ does not show significant correlations with $R'_d$ and $p'_{KS}$ (Kendall's Taus respectively are 0.214 and -0.048 for the confidence interval $0.262$), and weak negative correlations with $R_a$ and $\tau_K$ (Kendall's Taus respectively are -0.266 and -0.262 with the confidence interval $0.262$). Combining these results with the observation that the staying time does not correlate with the displacement at the population level, we infer that the effect of the dynamics on the staying time is rather unrelated to that on the displacement.

At last, we compare the empirical correlation with the modeling results reported in Ref. \cite{Han}. This model can create power-law distributed displacements from a random-walk process on a hierarchical geographical network. This series of displacements has inherent positive correlation due to the cascading process on the hierarchical organization. These modeling results and our empirical findings show high level of similarities for both the strength of the correlation and the pattern of its decay as the interval $m$ increases, implying that the cascading effect is the common origin of both the scaling displacement distribution and the positive correlation.
A more detailed discussion can be found in {\it Appendix D}.

\begin{table*}
\tabcolsep 0pt
\caption{
Information and fitting parameters of the 32 individuals, where $n_s$ and $n_d$ are the number of effective staying positions and displacements of each user, $d_*$ and $d_{max}$ are the average displacement and the maximum displacement of the user, and $\alpha$ is the fitting exponent of $P(d)$ using the estimated lower bound $x_{min}$. The definition of other parameters can be found in the main text.
} \vspace*{-10pt}
\scriptsize
\begin{center}
\def\temptablewidth{1.0\textwidth}
{\rule{\temptablewidth}{1pt}}
\begin{tabular*}{\temptablewidth}{@{\extracolsep{\fill}}ccccccccccccccc}
user ID & $n_s$ & $n_d$ & $d_*$/m & $d_{max}$/m & $\tau_K$ & $\Delta \tau$ & $R_a$ & $H$ & $p_{KS}$ & $x_{min}$ & $\alpha$ & $p'_{KS}$ & $R'_d$ & $R'_t$ \\ \hline
1 & 424 & 407 & 10980 & 879844 & 0.231 & 0.085  & 0.360  &0.83 & $4.37 \times 10^{-3}$ & 49.9  & 1.496  & $1.21\times10^{-17}$ & -0.978  & -0.989\\
2 & 278 & 257 & 12180 & 258183 & 0.360 & 0.108  & 0.500  &0.85 & $3.98\times 10^{-1}$ & 3203.3  & 1.604  & $1.05\times10^{-7}$ & -0.983  & -0.980\\
3 & 747 & 716 & 14311 & 5261960 & 0.312 & 0.065  & 0.421  &0.80& $6.42\times 10^{-1}$ & 404.9  & 1.469  & $2.22\times10^{-26}$ & -0.989  & -0.993\\
4 & 232 & 220 & 7301 & 257901 & 0.390 & 0.113  & 0.515  &0.66& $6.76\times 10^{-1}$ & 4477.9  & 2.084  & $3.97\times10^{-13}$ & -0.968  & -0.991\\
6 & 323 & 310 & 1373 & 40117 & 0.327 & 0.099  & 0.392  &0.79& $2.41\times 10^{-1}$ & 97.6  & 1.547  & $1.13\times10^{-5}$ & -0.989  & -0.974\\
8 & 237 & 228 & 905 & 21326 & 0.300 & 0.111  & 0.325  &0.55& $3.89\times 10^{-1}$ & 152.5  & 1.649  & $2.91\times10^{-4}$ & -0.980  & -0.991\\
9 & 1036 & 996 & 3161 & 490551 & 0.368 & 0.056  & 0.504  &0.84& $1.23\times 10^{-1}$ & 9.9  & 1.426  & $1.21\times10^{-1}$ & -0.994  & -0.993\\
10 & 563 & 541 & 2022 & 158233 & 0.367 & 0.075  & 0.457  &0.86& $5.18\times 10^{-1}$ & 98.6  & 1.543  & $5.64\times10^{-7}$ & -0.992  & -0.997\\
12 & 299 & 290 & 1086 & 21866 & 0.288 & 0.100  & 0.348  &0.51& $9.51\times 10^{-1}$ & 97.6  & 1.752  & $7.32\times10^{-18}$ & -0.966  & -0.993\\
15 & 240 & 218 & 47632 & 526428 & 0.187 & 0.116  & 0.227  &0.72& $3.43\times 10^{-1}$ & 35114.0  & 2.065  & $9.23\times10^{-13}$ & -0.976  & -0.980\\
22 & 1050 & 986 & 3055 & 60918 & 0.230 & 0.055  & 0.344  &0.69& $4.72\times 10^{-1}$ & 5601.9  & 3.232  & $1.10\times10^{-24}$ & -0.964  & -0.972\\
26 & 2702 & 2650 & 370 & 16074 & 0.376 & 0.034  & 0.546  &0.85& $3.62\times 10^{-1}$ & 15.2  & 1.656  & $3.07\times10^{-10}$ & -0.987  & -0.996\\
27 & 297 & 287 & 10732 & 639964 & 0.380 & 0.104  & 0.546  &1.00& $1.44\times 10^{-1}$ & 35.4  & 1.335  & $1.71\times10^{-3}$ & -0.985  & -0.986\\
28 & 729 & 681 & 5271 & 103931 & 0.326 & 0.068  & 0.403  &0.72& $1.74\times 10^{-7}$ & 10.1  & 1.277  & $1.82\times10^{-7}$ & -0.985  & -0.983\\
29 & 296 & 279 & 14295 & 867024 & 0.188 & 0.103  & 0.217  &0.70& $8.50\times 10^{-1}$ & 132.7  & 1.512  & $2.92\times10^{-17}$ & -0.984  & -0.986\\
34 & 243 & 233 & 1682 & 34774 & 0.443 & 0.110  & 0.554  &0.99& $8.35\times 10^{-1}$ & 264.9  & 1.692  & $3.16\times10^{-5}$ & -0.975  & -0.967\\
37 & 243 & 237 & 731 & 25306 & 0.231 & 0.110  & 0.307  &0.78& $9.61\times 10^{-1}$ & 21.7  & 1.692  & $2.88\times10^{-5}$ & -0.982  & -0.983\\
39 & 731 & 673 & 1138 & 6481 & 0.153 & 0.068  & 0.244  &0.53& $6.37\times 10^{-21}$ & 10.0  & 1.316  & $6.03\times10^{-21}$ & -0.898  & -0.959\\
40 & 379 & 365 & 1234 & 18140 & 0.518 & 0.093  & 0.655  &0.88& $9.56\times 10^{-1}$ & 2282.1  & 2.344  & $1.32\times10^{-11}$ & -0.963  & -0.996\\
41 & 234 & 224 & 1533 & 33011 & 0.373 & 0.114  & 0.471  &0.81& $1.06\times 10^{-1}$ & 116.9  & 1.498  & $4.49\times10^{-5}$ & -0.986  & -0.983\\
42 & 290 & 278 & 982  & 37454 & 0.363 & 0.101  & 0.446  &0.79& $7.77\times 10^{-1}$ & 225.8  & 1.706  & $2.24\times10^{-4}$ & -0.984  & -0.990\\
43 & 382 & 357 & 4419 & 72620 & 0.369 & 0.090  & 0.438  &0.91& $1.39\times 10^{-3}$ & 10.7  & 1.293  & $1.15\times10^{-3}$ & -0.987  & -0.989\\
44 & 361 & 354 & 212  & 9105 & 0.333 & 0.091  & 0.544  &0.88& $4.25\times 10^{-1}$ & 20.8  & 1.827  & $6.84\times10^{-5}$ & -0.979  & -0.982\\
46 & 215 & 204 & 2206 & 19166 & 0.260  & 0.119  & 0.333  &0.66& $2.58\times 10^{-1}$ & 73.2  & 1.443  & $1.44\times10^{-7}$ & -0.970  & -0.989\\
52 & 258 & 248 & 1497 & 39341 & 0.242  & 0.106  & 0.308  &0.54& $3.73\times 10^{-1}$ & 867.4  & 2.023  & $2.44\times10^{-8}$ & -0.981  & -0.992\\
54 & 823 & 796 & 8716 & 1159507 & 0.442  & 0.064  & 0.578  &0.82& $1.04\times 10^{-1}$ & 9.9  & 1.408  & $1.02\times10^{-1}$ & -0.992  & -0.993\\
78 & 319 & 306 & 4432  & 151888 & 0.416  & 0.096  & 0.509  &0.79& $8.21\times 10^{-1}$ & 17.5  & 1.427  & $2.63\times10^{-2}$ & -0.990  & -0.993\\
79 & 549 & 530 & 1038  & 78978  & 0.444  & 0.076  & 0.568  &0.81& $1.72\times 10^{-1}$ & 12.7  & 1.490  & $7.41\times10^{-3}$ & -0.991  & -0.985\\
116 & 272 & 261 & 567  & 13700  & 0.390  & 0.105  & 0.412  &0.82& $8.14\times 10^{-1}$ & 10.9  & 1.525  & $7.65\times10^{-1}$ & -0.986  & -0.986\\
123 & 247 & 234 & 6133  & 255818  & 0.437  & 0.114  & 0.576  &0.83& $9.76\times 10^{-1}$ & 3999.2  & 2.024  & $3.58\times10^{-13}$ & -0.970  & -0.994\\
134 & 360 & 340 & 9839  & 624928  & 0.348  & 0.094  & 0.416  &0.86& $1.01\times 10^{-1}$ & 47.5  & 1.339  & $5.37\times10^{-10}$ & -0.985  & -0.984\\
137 & 528 & 483 & 21151  & 1291274  & 0.374  & 0.079  & 0.479  &0.81& $6.66\times 10^{-3}$ & 92.6  & 1.304  & $8.13\times10^{-7}$ & -0.991  & -0.997\\
    \end{tabular*}
       {\rule{\temptablewidth}{1pt}}
       \end{center}
\end{table*}

\section{Conclusions and Discussions}

By analyzing the dataset of GPS carriers, we observe the scaling temporal-spatial properties in the aggregated human movements as well as individual-level diversities.
The displacement distribution at the population level is well-fitted by a power law. However, the individuals' mobility shows much diversity: some of them display common scaling properties, but others are irregular, in agreement with several recent studies \cite{Yan2}.


Our most remarkable finding is the significance positive correlation of the series of displacements both at the population level and at the individual level, showing that the gradually changing nature is wide-spread in human mobility. We surprisingly find that the strength of the correlation for each individual is significantly related with their displacement distribution: the individuals with stronger displacement correlation have a higher probability to possess a power-law-like displacement distribution. This result is confirmed by four types of correlations (Fig. \ref{indcorr}) and implies that the cascading-like dynamics is an important mechanism in the emergence of scaling properties of human mobility. Although the samples in our analysis are not very big, this result is still highly believable, as most of the correlations/correlations well pass the test with 95\% significant level and support each other.

We notice that the displacements and staying times are largely independent both at the population level and at the individual level, indicating that the mechanisms that drive the emergences of their scaling laws are also independent. This result is helpful for the modeling, as it indicates that we can divide the empirical findings into several classes that may have similar dynamics according to their correlations, and then can be modeled independently.

Finally, the speed-displacement pattern shows the abnormal transition from a linear to a sub-linear relationship (Fig. \ref{speed}), which may indicate the change of transportation from walks to automobile and the average longest walking distance in daily life.  In addition, the impact on human mobility patterns due to traffic jams are observed here.

In summary, we find that the positive correlation of the series of displacements that describes the cascading-like movements, is a characteristic factor that is relevant to the underlying mechanisms of the scaling of mobility patterns from the empirical analysis.
Our findings and the methods used provide some useful insights for further empirical and modeling studies of human mobility patterns.

\begin{acknowledgments}

This work was supported by the National Natural Science Foundation of China Grants Nos. 11205040, 11105024, 70971089 and 10635040. XPH acknowledges the Zhejiang Provincial Natural Science Foundation of China (Grant No. LY12A05003), the start-up foundation and Pandeng project of Hangzhou Normal University.
We thank Dr. M. Pleimling for his helpful discussions on this paper.

\end{acknowledgments}

\appendix

\section{Dataset descriptions and the judgement of effective stay positions}

\begin{figure}
  \includegraphics[width=8.5cm]{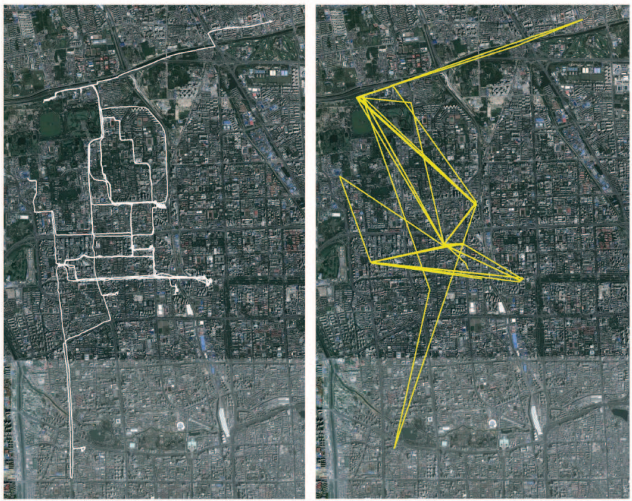}\\
  \caption{(Color online) Snapshot of the distinguishing of effective staying positions. The figure left shows original trajectory of one GPS carrier. The figure right shows the effective staying positions connected by lines in order, where each vertex represents an effective staying position.}\label{state}
\end{figure}

The data used in this study has been provided by the Microsoft Geo-life project and contains over 2 years of GPS trajectories (from April 2007 to August 2009) of 165 individuals. The datasets are available at the website: http://research.microsoft.com/en-us/downloads /b16d359d-d164-469e-9fd4-daa38f2b2e13/.
The GPS data was collected by different GPS handheld equipments or GPS phones. In most of them, the interval of recording time ranges from 2 to 5 seconds. The data set includes more than 10,000 trajectories, the total recording distance is more than 1 million kilometers, and the total recording time is more than 48,000 hours. The trajectories are widely distributed in the world, covering more than 30 cities in China, and several cities in North America, Europe, South-east Asia, etc. The movements recorded by the data set include not only trips to work or home, but also many daily-life activities, such as shopping, sightseeing, dining, hiking, and cycling, etc. The recording time for different individuals is different, and ranges from several weeks to several years. A trajectory file consists of a sequence of the records of trajectory points, and each record provides information on the latitude, longitude, and altitude of the position of the GPS holder, and the corresponding recording time.

The data sets are composed by a series of geographic locations with corresponding time recordings ordered by the time sequence. They can not directly show the positions that users really have stayed in, so first of all we should distinguish the effective stay positions from the record.
We set the resolutions for distinguishing of staying positions to 10 meters for the displacement which is the critical spatial resolution of a handheld GPS equipment, and 120 seconds for the time which is the interval of traffic signals.

Consider a trajectory labelled by $\{P_1,P_2,\cdots,P_N\}$, where a continuous sub-sequence $\{P_j,\cdots,P_k\}$ (where $1\leq j\leq k\leq N$) satisfies the following two conditions: the distances between two consecutive track points are less than 10 meters, and the total time length of the sub-sequence $\{T_j,\cdots,T_k\}$ is larger than 120 seconds. The average position of the sub-sequence is recorded as an effective stay position, and $t_s = T_k - T_j$ is the staying time of the stay position. As illustrated in Fig. 1, the average position S1 of track points from P3 to P7 are considered as an effective stay point,as all the geographical distances from P3 to P7 are no more than 10m and $T_7 - T_3 < 120s$. The same holds true for S2 for the track points from P10 to P15. The straight-line distance between S1 to S2 is set as the user's displacement for the movement from S1 to S2.

Most of the files in the data set only contain the records of few hours or minutes. Since the critical staying time in each stay position is set as 120 seconds, we usually can not obtain enough effective stay positions to achieve good patterns of user's mobility, and we therefore abandon all the files where the recording time is less than 6 hours and , we are left with 927 files from 100 users. Using the above algorithms, we distinguish the effective stay positions of each of the 100 users from the 927 files, which are used in our analysis at the population level.

However, in our empirical analysis at the individual level, the number of effective stay positions of more than half of the 100 users is too small to extract its patterns. We thus remain with the data of 32 users with a number of effective stay positions that is larger than 200.
Notice that we analyze the files of a same user one by one, and the statistical patterns of the user is aggregated from all of his/her files.

\section{Kendall's Tau}

In our empirical analysis, the displacements of the users are very heterogeneous, covering several orders of magnitude.  Thus classical measurements like the Pearson coefficient are not suitable in analyzing the correlation of these displacements. We therefore use the rank-based correlation coefficient named \emph{Kendall's Tau}.
For two series $\vec{x}=\{x_1,x_2,\cdots,x_m\}$ and $\vec{y}=\{y_1,y_2,\cdots,y_m\}$, the Kendall's Tau is
defined as \cite{Kendall1938}
\begin{equation}
\tau_K = \frac{2}{m(m-1)}\sum_{i<j}\texttt{sgn}[(x_i-x_j)(y_i-y_j)],
\end{equation}
where $\texttt{sgn}(x)$ is the signum function, which equals +1 if $x>0$, -1 if $x<0$, and 0 if $x=0$. $\tau_K$ ranges from +1 (exactly the same ordering of $\vec{x}$ and $\vec{y}$) to -1 (reverse ordering of $\vec{x}$ and $\vec{y}$), and two uncorrelated series have $\tau_K\approx 0$.
Obviously, as $\tau_K$ is calculated based on the order of the elements in two series, the magnitudes of differences on the value of the elements do not impact $\tau_K$.

\section{Detrended fluctuation analysis}

The detrended fluctuation analysis (DFA) is a method proposed to evaluate the self-affinity of a time series in stochastic processes. It was first developed by Peng, et al. \cite{DFA}, and is helpful to reveal the extent of long-term correlations of a time series. Using the DFA method, the Hurst exponent can be derived through the following procedures.

i) Considering a time series \{$d_t, t \in \mathbb{N} $\}, we first need to calculate the integration $D(t)$ of the time series,
\begin{equation}
D(t) = \sum_{i=1}^{t}(d_{i}-<d_t>).
\end{equation}
where $<\cdot>$ means an average over all $t's$.

ii) Then divide $D(t)$ into mutually disjoint boxes of size $\Delta t$.

iii) In each box, using the least square method, we can get a $n$-order polynomial fit $D_{fit}(t)$, which is called the $n$-order trend. The residual series, in which the trend has been eliminated, can be derived by applying a subtraction.
\begin{equation}
Y(t)=D(t)-D_{fit}(t).
\end{equation}

iv) Calculate the mean square error of each box over the size $\Delta t$ after eliminating the trend.
\begin{equation}
E(\Delta t)^2 = \sum_{t=1}^{\Delta t} Y(t)^2,
\end{equation}

v) Next calculate the root-mean-square deviation, or say fluctuation, over different $(\Delta t)'s$.
\begin{equation}
F(\Delta t) = \sqrt{\displaystyle \frac{1}{\Delta t} E(\Delta t)^2}.
\end{equation}

vi) If the time series $\left\lbrace d_t \right\rbrace$ satisfies a power-law distribution, the quality $F(\Delta t)$ will also follow a power-law increasing function,
\begin{equation}
F(\Delta t) \sim (\Delta t)^H,
\end{equation}
where $H$ is the Hurst exponent that we want to calculate.
Here $H=0.5$ represents the time series is completely uncorrelated, and $0.5<H<1.0$ indicates the time series is of long-term correlation.

\section{Comparison with the purely cascading-like mobility process}

\begin{figure}
  \includegraphics[width=8.7cm]{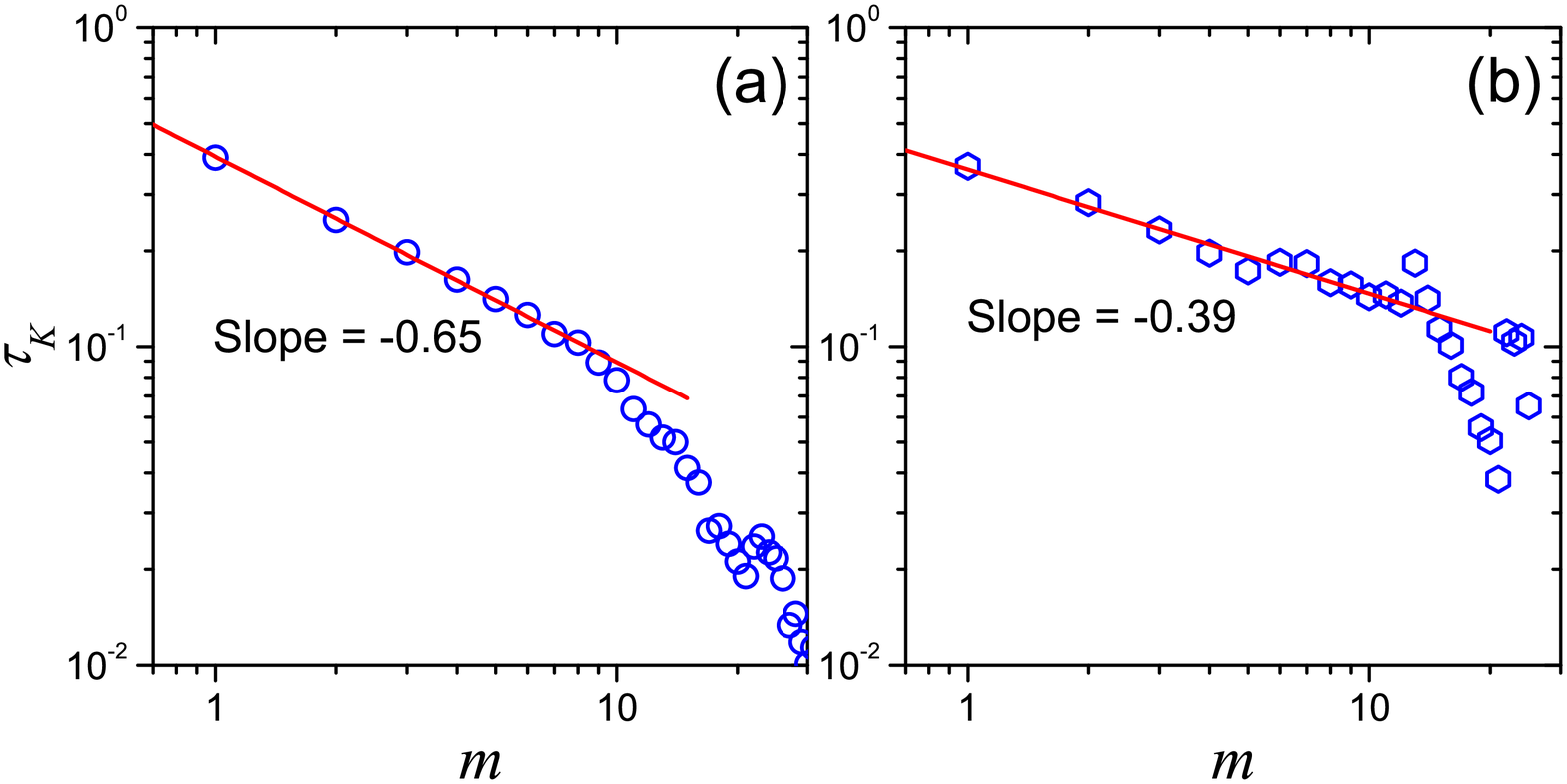}\\
  \caption{(Color online)
  The decay of Kendall's Tau $\tau_K$ between $\vec{d}_i$ and $\vec{d}_{i+m}$ as a function of the interval $m$ for modeling series (a) and No. 9 user (b).
  }\label{model}
\end{figure}

Our findings show the significant positive correlation among the displacements of human mobility. It is necessary to compare these empirical results with the modeling displacement series with inherent positive correlation and scaling properties. Here the hierarchical-traffic-system model reported in Ref. \cite{Han} is considered.

The basic rules and results of the model can be briefly introduced as follows:
Firstly we create a hierarchical geographic network on a two-dimensional plane. In the plane, $K$ top-layer nodes, $K(M-1)$ 2nd-layer nodes, $\cdots$, $KM^{n-2}(M-1)$ $n$th-layer ($n>2$) nodes, $\cdots$, and $KM^{N-2}(M-1)$ $N$th-layer nodes are randomly distributed on the plane. Each node is then connected to its nearest up-layer node. For the $n$th-layer node, its weight is $w_n = r^{N-n}$, where $r>1$ presents the upper layer nodes that have more attraction for agents. After the construction of the hierarchical network, an agent randomly walks on it. The probability that the agent will move to a neighboring city is proportional to its weight.
Simulations and analytical results show that the agent's displacement distribution obeys power law with exponent $\beta = 3-4\log_Mr$. When $M = 9$ and $r = 2$, $\beta \approx 1.7$, which is very close to the empirical findings.

Obviously, due to the hierarchical organization, the probability that walkers directly move from a top-layer node to a bottom layer node is small. Since the long range movements only appears between two higher layer nodes, the displacement of the agents is gradual changed and has inherent positive correlation. We calculate the Kendall's Tau of the displacement series, and its value is 0.39, very close to our empirical results. In addition, $\tau_K$ of the modeling series also shows the power law decay when $m$ increases, as shown in Fig. \ref{model}(a), which is comparable with Fig. \ref{Tau}.

\end{document}